\documentclass[11pt]{article}

\pdfoutput=1

\usepackage{booktabs} 

\usepackage[normalem]{ulem}

\usepackage{float}
\usepackage{epsfig}
\usepackage{cite}
\usepackage{amsmath}
\usepackage[footnotesize]{caption}
\usepackage{slashed}
\usepackage{xcolor}
\usepackage[utf8]{inputenc}
\usepackage{tikz}
\usetikzlibrary{shapes.geometric}
\usetikzlibrary{arrows.meta,arrows}
\usetikzlibrary{quotes,angles}

\usepackage{multicol}
\usepackage{multirow}
\usepackage{array}

\numberwithin{equation}{section}

\newcommand{\be}{\begin{equation}}
\newcommand{\ee}{\end{equation}}
\newcommand{\beq}{\begin{eqnarray}}
\newcommand{\eeq}{\end{eqnarray}}

\usepackage{bm}

\begin{document}

\title{
\vspace*{-3.0cm}
\phantom{h} \hfill\mbox{\small }\vspace*{-1.1cm}
\vspace*{0.7cm}
\\[1cm]
\vspace{13mm}
\textbf{Search for an invisible scalar in $t \bar{t}$ final states\\ at the LHC\\[4mm]}}

\date{}
\author{
Duarte Azevedo$^{1,2\,}$\footnote{E-mail:
  \texttt{duarte.azevedo@kit.edu}} ,
Rodrigo Capucha$^{3\,}$\footnote{E-mail:
\texttt{rscapucha@fc.ul.pt}} ,
Pedro Chaves$^{4\,}$\footnote{E-mail:
\texttt{pshalo06@gmail.com}} ,\\
Jo\~ao Bravo Martins$^{3\,}$\footnote{E-mail:
\texttt{joaobravomartins24@hotmail.com}} , 
Ant\'{o}nio Onofre$^{4\,}$\footnote{E-mail:
\texttt{antonio.onofre@cern.ch}},
Rui Santos$^{3,5\,}$\footnote{E-mail:
  \texttt{rasantos@fc.ul.pt}} 
\\[5mm]
{\small\it
$^1$Institute for Theoretical Physics, Karlsruhe Institute of Technology,} \\
{\small\it Wolfgang-Gaede-Str.~1,
76128 Karlsruhe, Germany}\\[3mm]
{\small\it
$^2$Institute for Astroparticle Physics, Karlsruhe Institute of Technology,} \\
{\small\it Hermann-von-Helmholtz-Platz 1, 76344 Eggenstein-Leopoldshafen, Germany}\\[3mm]
{\small\it $^3$Centro de F\'{\i}sica Te\'{o}rica e Computacional,
    Faculdade de Ci\^{e}ncias,} \\
{\small \it    Universidade de Lisboa, Campo Grande, Edif\'{\i}cio C8
  1749-016 Lisboa, Portugal} \\[3mm]
{\small\it
$^4$ Centro de F\'{\i}sica da Universidade do Minho e Universidade do Porto (CF-UM-UP),}\\ {\small\it Universidade do Minho, 4710-057 Braga, Portugal} \\[3mm]
{\small\it
$^5$ISEL -
 Instituto Superior de Engenharia de Lisboa,} \\
{\small \it   Instituto Polit\'ecnico de Lisboa
 1959-007 Lisboa, Portugal} \\[3mm]
}

\maketitle

\begin{abstract}
\noindent

We use the current $t\bar t$ experimental analysis to look for Dark Matter (DM) particles hidden in the final state.
 We present a phenomenological study where we successfully perform the reconstruction of a $t\bar{t}$ system in the presence of a scalar mediator $Y_0$, that couples to both Standard Model (SM) and to DM particles. We use a \texttt{MadGraph5\_aMC@NLO} simplified DM model, where signal samples of $pp \rightarrow t\bar{t}Y_0$ are generated at the Large Hadron Collider (LHC) with both Charge-Parity (CP) -even and CP-odd couplings of $Y_0$ to the top quarks. Different mass scales for the $Y_0$ mediator are considered, from the low mass region ($\sim$ 0~GeV) to masses close to the Higgs boson mass (125~GeV).  
The dileptonic final states of the $t\bar{t}$ system were used in our analysis. The reconstruction of the $t\bar{t}$ system is done with a kinematic fit, without reconstructing the mediator. All relevant SM backgrounds for the dileptonic $t\bar{t}$ search at the LHC are considered. Furthermore, CP angular observables were used to probe the CP-nature of the coupling between the mediator and top-quarks, which allowed to set confidence level (CL) limits for those Yukawa couplings as a function of the mediator mass. 

\end{abstract}

\thispagestyle{empty}
\vfill
\newpage
\setcounter{page}{1}

\section{Introduction}
 
Evidence for the existence of dark matter (DM) provided by astrophysical observations, be it from gravitational lensing effects \cite{hoekstra2002nf,koopmans2002qh,metcalf2003sz,moustakas2002iz}, galactic rotational velocity curves \cite{rubin1970rotation} or from other measurements like the collision of clusters of galaxies (Bullet Cluster) \cite{clowe2004weak}, is by now overwhelming. However, the nature of this non-luminous matter remains unknown in spite of decades of intensive searches in a vast array of experiments. From the observations, we can hypothesise that if DM is actually composed of particles, they only interact very weakly with the Standard Model (SM). Since SM neutrinos have been mostly ruled out as possible DM candidates, due to cosmological constraints \cite{weinheimer2003laboratory}, this suggests the possible existence of a "dark sector" composed of particles beyond the SM. Attempts at direct and indirect detection of the myriad of particles proposed from extending the SM with such a sector (such as weakly interacting massive particles and axions, among many others) have, so far, provided no definitive results favouring any particular hypotheses~\cite{rosenberg2000searches,schumann2019direct}.

DM searches at colliders have focused on mono-events, such as mono-jet, mono-Higgs among others~\cite{GonzalezFernandez:2021nbb, Kumar:2022knu, ATLAS:2023tkt}. 
There are also bounds on the portal couplings from the invisible branching ratio of the 125 GeV Higgs~\cite{Biekotter:2022ckj, Milosevic:2021bpv}.  
Searches for DM particles produced alongside a $t \bar t$ pair and single top quark events have also been performed in the past~\cite{Haisch:2016gry, Haisch:2018bby, Hermann:2021xvs, Haisch:2021ugv} mostly
using variables such as missing energy as a discriminator,  with no attempt to reconstruct the kinematics of the top quarks.
However, and to best of our knowledge, this is the first time a study is designed to look for DM particles hidden in the current on-going searches and analysis. 
Let us consider as an example the case of $t \bar t$ production in the di-leptonic channel. 
The question we want to answer is: if a very light DM particle would be produced alongside the $t \bar t$ pair would we see any difference in any distributions? 
If no differences are found, could the analysis already performed be used to limit the couplings of DM particles for very light invisible particles? One should note
that from the point of view of high energy collider physics we are really exploring the case of a zero DM mass particle which means that we are spanning 
DM masses from the scenarios of ultra-light particles to a few GeV in mass. That is, for LHC energies, masses below about 1 GeV are for all practical purposes equal to zero.
Still we will also show how the $t\bar t$ analysis performs in the case of a 125 GeV scalar.

In order to implement our idea we consider a new mediator that couples to both DM and  to the SM particles. We study the possibility of using previous mainstream experimental analysis of $pp \to t \bar t$ in the di-leptonic channel to
gauge the impact of  a spin-0 DM mediator ($ J^{CP}= 0^{\pm}$) in the associated production process $t\bar{t}Y_0$. The analysis is performed within the description of simplified models of DM production at the LHC, 
where the DM mediator ($Y_0$) couples to the top-quarks proportionally to the top mass. The results are presented as a function of the modifier of this new Yukawa coupling.
This is a convenient approach, as the LHC can explore a large spectrum of DM mediator masses and coupling strengths, allowing to access the CP-nature of these mediators, in case they exist even as mixed CP states.

This paper is organised as follows: the simplified DM model, the relevant parameters and the angular observables we used, are presented in Section~\ref{sec:TH}. The event generation and simulation are described in Section~\ref{sec:generation} and, in Section~\ref{sec:matching}, the event selection and kinematic reconstruction are discussed. Our results are presented in Section~\ref{sec:results} and the main conclusions are described in Section~\ref{sec:conclusions}.
\newpage

\section{The DM Lagrangian \label{sec:TH}}
\hspace{\parindent} 

In our study, we used the simplified DM model \texttt{DMsimp}~\cite{kentarou2015higher} where, besides the scalar $Y_0$ boson, we also have a dark sector that couples only to  $Y_0$ . In our paper, we will remain agnostic to the latter, focusing solely on the interaction between the $Y_0$ DM mediator to the SM content. In particular, we will assume Yukawa couplings proportional to the mass of the respective SM particle and hence dedicate ourselves exclusively to top quarks. The Lagrangian density can thus be simplified and written as follows
\begin{equation}
    \mathcal{L}_{SM}^{Y_0} = \frac{y_{33}^t}{\sqrt{2}} \bar{t}(g_{u_{33}}^S + ig_{u_{33}}^P \gamma^5) t Y_0 \quad,
\label{eq:SMinteraction}
\end{equation}
%
where the $g^{S/P}_{u_{33}}$ are the CP-even/-odd couplings of the DM mediator ($Y_0$) to top quarks, respectively. They are normalized to the SM Yukawa couplings, $y^f_{ii}=\sqrt{2}m_f/v$. The scalar hypothesis (CP=+1) is given by setting $g^S_{u_{33}} = 1$ and $g^P_{u_{33}} = 0$ and for the pseudo-scalar scenario (CP=-1) we set $g^S_{u_{33}} = 0$ and $g^P_{u_{33}} = 1$. 
When both $g^{S/P}_{u_{33}} \neq 0$, the interaction has both CP-even and -odd components and is thus CP-violating.

Our goal is to explore how angular distributions of SM particles may help probing the dark sector, by looking into the expected changes of these observables in the presence of this DM mediator. Several CP-observables have been proposed in the literature to probe the CP-nature of the coupling of top quarks to the Higgs boson at the LHC or future colliders, using mainly the $t\bar{t}H$ channel~\cite{Bernreuther_1994, Gunion_1996, Bhupal_Dev_2008, Frederix_2011, Ellis_2014, Khatibi_2014, Demartin_2014, Kobakhidze_2014, Bramante_2014, Boudjema_2015, He_2015, Amor_dos_Santos_2015, Gritsan_2016, Dolan_2016, Gon_alves_2016,  Buckley_2016_v2, Mileo_2016, Buckley_2016, Amor_dos_Santos_2017, Gon_alves_2018, Azevedo_2018, Li_2018, Ferroglia_2019, Faroughy_2020, Azevedo:2020vfw, Azevedo:2020fdl, Bortolato_2021, new_obs, Gon_alves_2022, Barman_2022, Azevedo_2022}. 
The vast majority of these variables are only sensitive to the square terms $(g^S_{u_{33}})^2$ and $(g^P_{u_{33}})^2$ that appear in the cross section of the interaction described by equation~\ref{eq:SMinteraction}.
%

After looking in detail at several possible observables, we concluded that the most effective ones are the azimuthal angle difference $\Delta \phi_{\ell^-\ell^+}$ of the charged leptons that come from the decay of top quarks, and the $b_4$ variable evaluated in the laboratory frame (LAB), $b_4 = (p^z_t . p^z_{\bar{t}}) / (|\vec{p}_{t}| . |\vec{p}_{\bar{t}}| )$,
where the $z$-direction corresponds to the beam direction, and $\vec{p}_{t(\bar{t})}$ and $p^z_{t(\bar{t})}$ correspond to the total and z-component of the top (anti-top) quark momentum measured in the LAB frame, respectively. It is worth noting that the $b_4$ variable depends on the $t$ and $\bar{t}$ polar angles, $\theta_t$ and $\theta_{\bar{t}}$ respectively, with respect to the $z$-direction, and can be expressed as $b_4=\cos{\theta_t} \times \cos{\theta_{\bar{t}}}$. In order to evaluate this variable, the kinematic reconstruction of the $t\bar{t}$ system needs to be accomplished. This will be discussed in the next sections.

\section{Event generation and simulation
\label{sec:generation}}
\hspace{\parindent} 

\noindent
LHC-like signal and background events were generated with \texttt{MadGraph5\_aMC@NLO}~\cite{Alwall:2011uj}, with a centre-of-mass energy of 13~TeV. The DM simplified model, \texttt{DMsimp}~\cite{kentarou2015higher}\footnote{Available in the \texttt{FeynRules} repository.}, was used to generate $pp\to t\bar tY_0$ signal events, at Leading Order (LO). The pure scalar and pseudo-scalar signals were generated by setting the respective couplings as mentioned in the previous section (following the Lagrangian density in equation~\ref{eq:SMinteraction}). The mass of the DM mediator was set to $m_{Y_{0}}=0,1,10$ and $125$~GeV, and the top quark mass to $m_t=172.5~GeV$. Only the dileptonic final state of the $t\bar{t}$ system was considered ($t\bar{t}\rightarrow bW^+\bar{b}W^-\rightarrow b\ell^+\nu_\ell\bar{b}\ell^-\bar{\nu}_{\ell}$), with the DM mediator forced not to decay, although if we allow the DM mediator to decay to mostly DM particles, the analysis and subsequent results do not change. Backgrounds from SM $t\bar{t}$ (plus up to 3~jets), $t\bar{t}V$ (plus up to 1 jet), $t\bar{t}H$, single top quark production ($t$-, $s$- and $Wt$-channels), $W/Z$ (plus up to 4~jets), $W$($Z$)$b\bar{b}$ (plus up to 2~jets) and $WW, ZZ, WZ$ diboson processes were also generated using \texttt{MadGraph5\_aMC@NLO} at LO. 
Following event generation and hadronization by \texttt{PYTHIA}~\cite{Sjostrand:2006za}, all signal and background events went through a fast simulation of a typical LHC detector performed by \texttt{Delphes}~\cite{deFavereau:2013fsa}, using the default ATLAS detector cards. Further details on the event generation and detector simulation can be found in~\cite{Azevedo:2020vfw}. The analysis of signal and background events is performed within the \texttt{MadAnalysis5}~\cite{Conte:2012fm} framework.\\  

\section{Event selection and kinematic reconstruction
\label{sec:matching}}
\hspace{\parindent} 



Events are selected by requiring the jets and leptons reconstructed by \texttt{Delphes} to have their pseudo-rapidity\footnote{The pseudo-rapidity is defined by $\eta=-ln[tan(\theta/2)]$, where $\theta$ is the particle polar angle.} $\eta<2.5$ and transverse momenta $p_T>20$~GeV. Only events with two jets and two isolated leptons of opposite charge are accepted. To avoid contamination from the $Z$ + jets background, we require the invariant mass of the two lepton system to fulfil $|m_{\ell^+ \ell^-}-m_Z|>10$~GeV. Further details on the event selection criteria can be found in~\cite{Azevedo:2020fdl}. 

In the reconstruction of the $t\bar{t}$ system, we assume the reconstructed leptons are the ones
from the W decays, originated in the top quark decays. We then need to assign the reconstructed jets to the correct b-quarks from the associated top quarks. In order to avoid mismatches, once two jets 
 are present in the final state, we use multivariate statistical methods from the Toolkit for Multivariate Data Analysis,  \texttt{TMVA}~\cite{hoecker2007tmva}, to find the right pairing of jets and leptons..
\begin{figure}[h!]
	\begin{center}
		\includegraphics[width = 7.8cm]{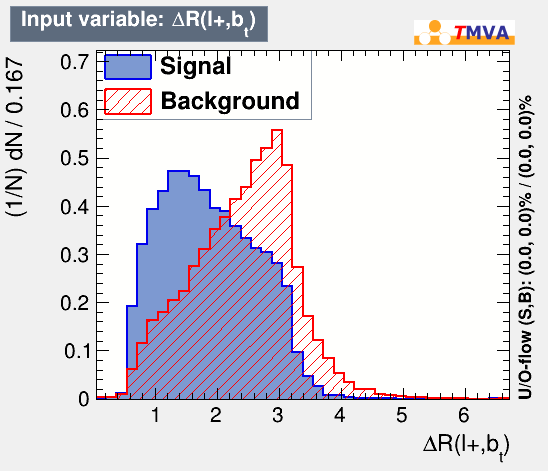}
		\includegraphics[width = 7.8cm]{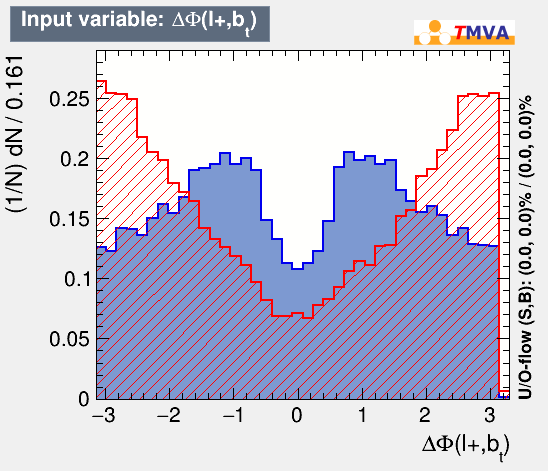}
		\caption{Normalized TMVA input variable distributions for correct combinations (labeled as {\it signal}, in blue) and wrong combinations (labeled as {\it background}, in red), for a DM $J^P=0^+$ mediator. The $\Delta R$ between the $\ell^+$ and the $b$-jet from the $t$ decay, is shown on the left plot. The corresponding $\Delta\Phi$ distribution can be seen on the right.}
		\label{fig:TMVAvars0plus}
	\end{center}
\end{figure} 
\begin{figure}[h!]
	\begin{center}
		\includegraphics[width = 7.8cm]{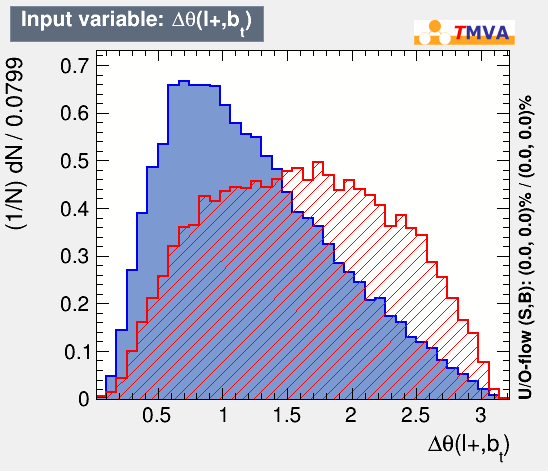}
		\includegraphics[width = 7.8cm]{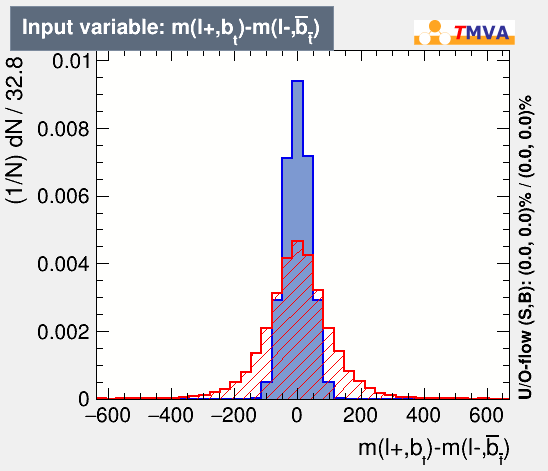}
		\caption{Normalized TMVA input variable distributions for correct combinations (labeled as {\it signal}, in blue) and wrong combinations (labeled as {\it background}, in red), for a DM $J^P=0^+$ mediator. The $\Delta \theta$ between the $\ell^+$ and the $b$-jet from the $t$ decay is shown on the left and the lepton plus $b$-jet mass difference is shown on the right.}
		\label{fig:TMVAvars0plus2}
	\end{center}
\end{figure} 

\begin{figure}[h]
	\begin{center}
		\includegraphics[width = 7.5cm]{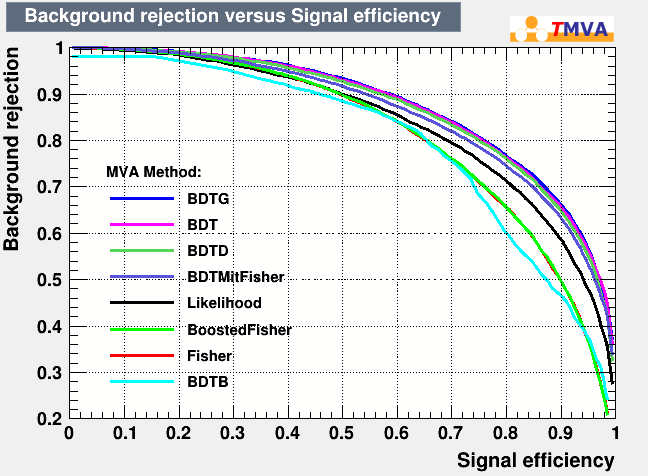}
		\includegraphics[width = 7.5cm]{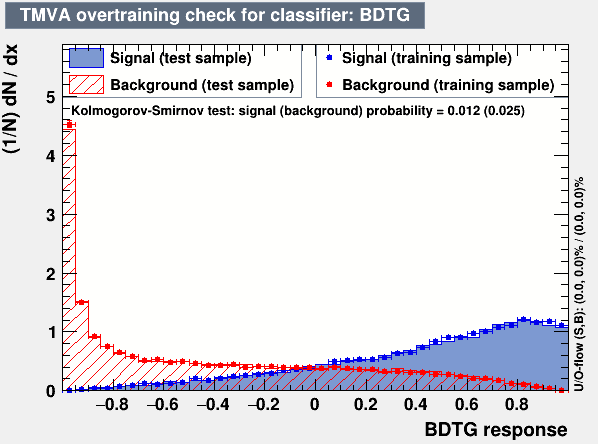} \\
		\includegraphics[width = 7.5cm]{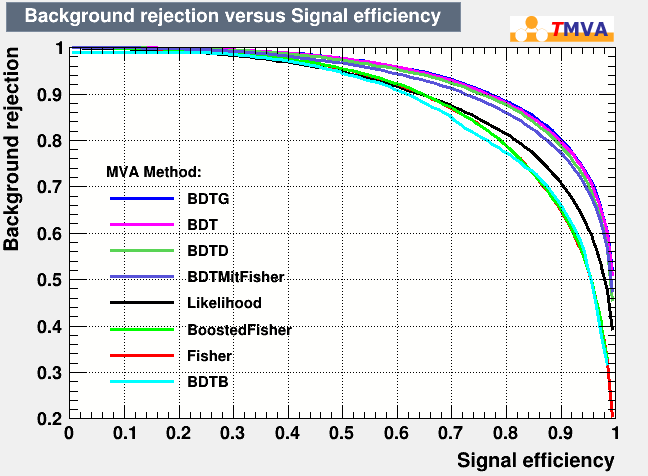}
		\includegraphics[width = 7.5cm]{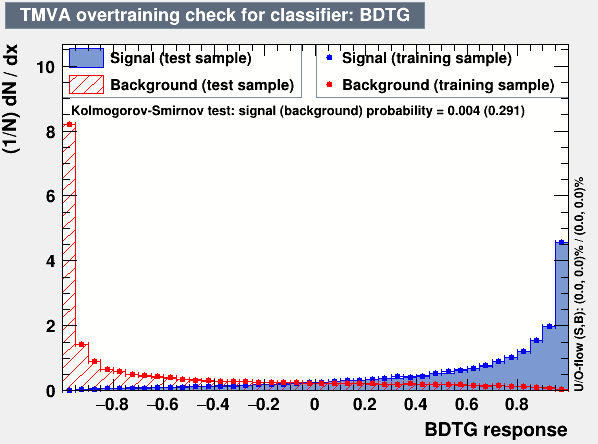} \\
		\caption{Background rejection versus signal acceptance (ROC curve) for different multi-variate methods are compared, for the $J^P=0^+$ mediator (top left). The distribution of the best classifier (BDTG) is also shown (top right). The equivalent plots for the case of the $J^P=0^-$ mediator, are shown in the bottom plots.}
		\label{fig:TMVAROC0plus}
	\end{center}
\end{figure}

To that effect we use several distributions (which already include hadronization and shower effects), where {\it right} and {\it wrong} combinations
are compared. A wrong combination happens whenever a jet not originating from a top decay is assigned to its corresponding b-quark.   
Right (labelled as Signal, in blue) and wrong (labelled as Background, in red) combinations are represented in Figures~\ref{fig:TMVAvars0plus} and~\ref{fig:TMVAvars0plus2}, for the $\Delta R$~\footnote{$\Delta R\equiv\sqrt{\Delta \Phi^2+\Delta \eta^2}$, where $\Delta \Phi \, (\Delta \eta)$ correspond to the difference in the azimuthal angle (pseudo-rapidity) of two objects.}, $\Delta\Phi$ and $\Delta\theta$ between the $\ell^+$ lepton and the jet from the hadronization of the $b$-quark (originated in the $t$ decay and labelled as $b_t$). 
Similar distributions were also obtained for the $\ell^-$ lepton, and used to optimize finding the right combination. 
Clear differences between the wrong and right combinations are visible in all distributions shown. The invariant mass difference between the pairs ($b_t, l^+$) and ($\bar{b}_{\bar{t}}, l^-$) is also shown in Figure~\ref{fig:TMVAvars0plus2}, for the right and wrong combinations. 
Several multivariate statistical methods were then trained by \texttt{TMVA}, using right and wrong combinations distributions for training and testing. All individual distributions were combined into a single discriminant classifier for each one of the methods. In Figure~\ref{fig:TMVAROC0plus} (left), we show the Receiver Operating Characteristic (ROC) curve for the $Y_0$ scalar ($J^P=0^+$) and pseudo-scalar ($J^P=0^-$) mediators in the top and bottom plots, respectively. From the ROC curves we can see that the best method is a Boosted Decision Tree with Gradient boost (BDTG).
The corresponding classifier outputs are shown in Figure~\ref{fig:TMVAROC0plus} (right), for scalar and pseudo-scalar DM mediators in the top and bottom plots, respectively.
From the comparison between both cases, we can see that the scalar mediator is more challenging, with a slightly worse ROC curve. For this reason, from this point on, all results shown originate directly from the scalar mediator analysis, since this case represents the most conservative scenario.

To reconstruct the 3-momentum of the undetected neutrinos from the top quark decays, we impose the following energy-momentum conservation conditions to events,
\begin{align}
    (p&_{\nu} + p_{\ell^+})^2 = m^2_{W} , \nonumber \\
    (p&_{\bar{\nu}} + p_{\ell^-})^2 = m^2_{W} , \nonumber \\
    (p&_{W^+} + p_b)^2 = m^2_t , \label{eq:etmiss} \\ 
    (p&_{W^-} + p_{\bar{b}})^2 = m^2_{\bar{t}} , \nonumber \\
    p&^x_{\nu} + p^x_{\bar{\nu}} = \slashed{E}^x , \nonumber \\
    p&^y_{\nu} + p^y_{\bar{\nu}} = \slashed{E}^y , \nonumber 
\end{align}

\noindent
where $p_\varsigma$ ($p_\varsigma^\kappa$) represents the four-momentum of particle $\varsigma$ (its projection along the $\kappa$-axis).
In the first four equations mass constraints are imposed, where neutrinos and charged leptons are assumed to reconstruct the masses of the $W$ bosons they originated from which, when combined with the right jet, should reconstruct the correspondent top quark masses. We also consider (last two equations) the total missing transverse energy is wholly accounted for by the neutrinos. In this approximation, we are assuming the DM mediator contribution to the missing transverse energy to be negligible (as well as its $z$-axis component) when compared to the neutrinos contribution. In order to find the best solution for each event, the top quark and $W$ boson mass values, used by the fit, were sampled 500 times from 2-dimensional probability distribution functions (\textit{p.d.f.}s) obtained from parton level (with shower effects) $t\bar{t}Y_0$ signal events (see~\cite{Azevedo:2020fdl}, for more details). 
\begin{figure}[H]
	\begin{center}
		\includegraphics[width = 7.0cm]{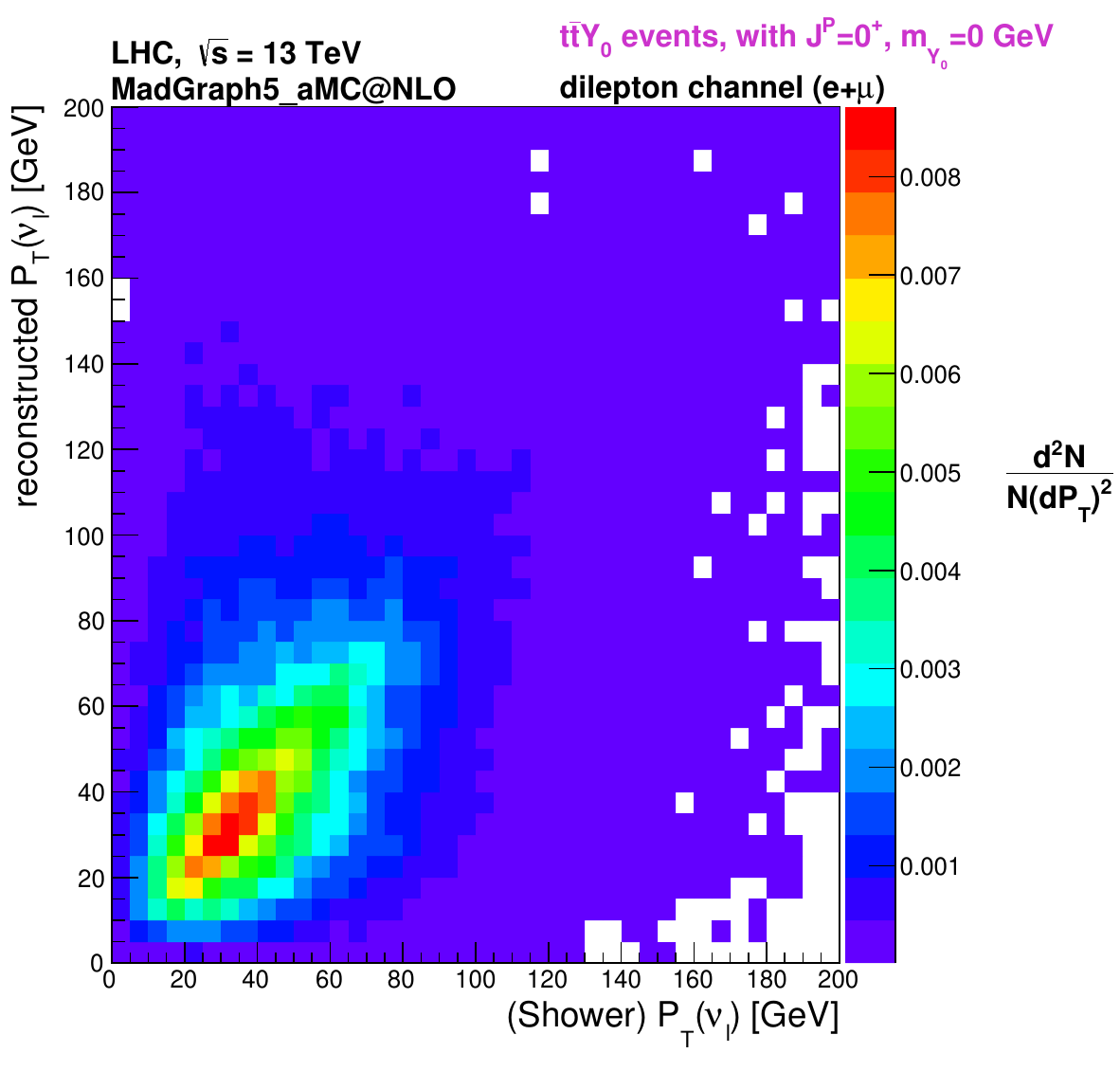}
		\includegraphics[width = 7.0cm]{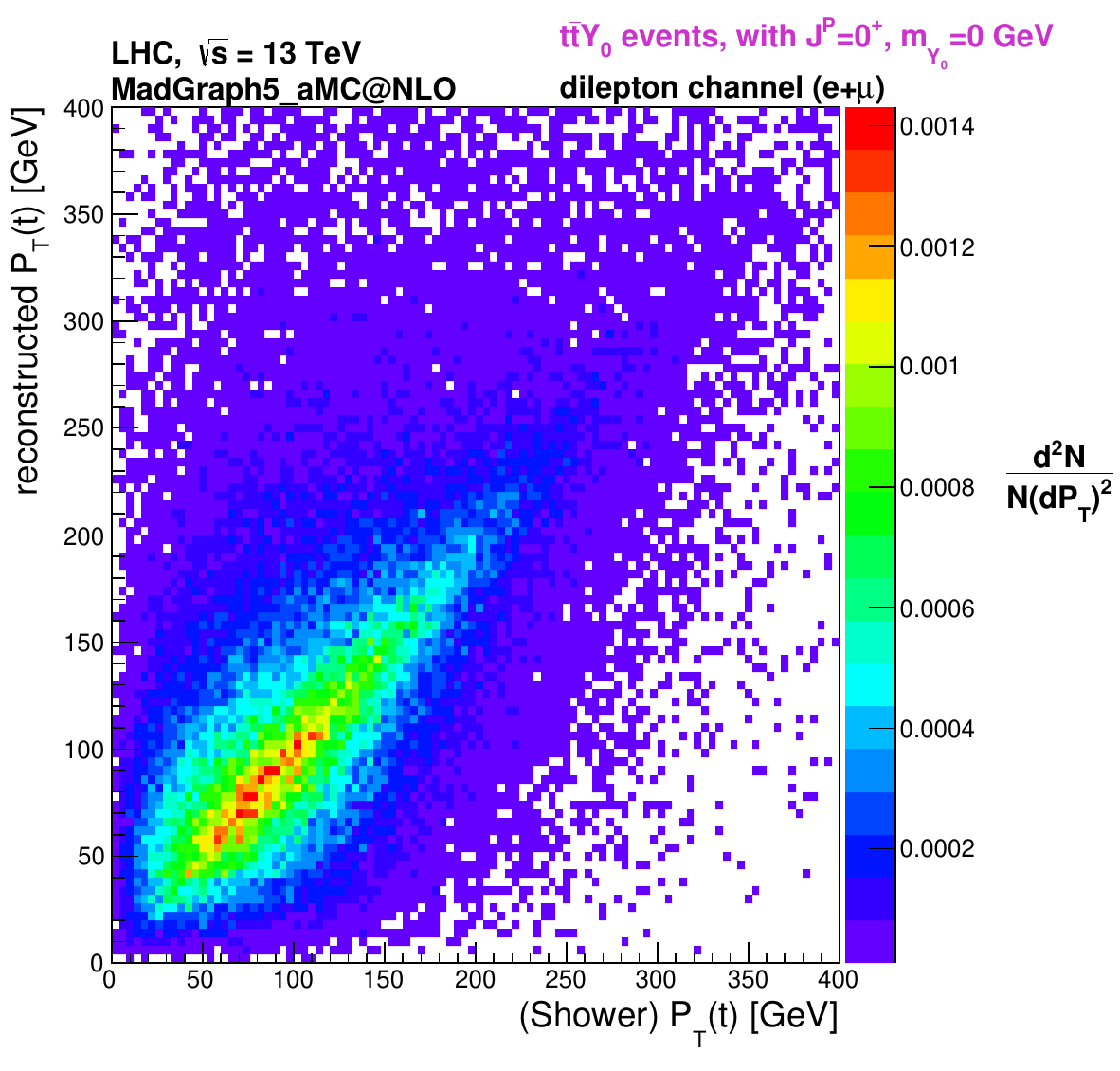} \\
		\includegraphics[width = 7.0cm]{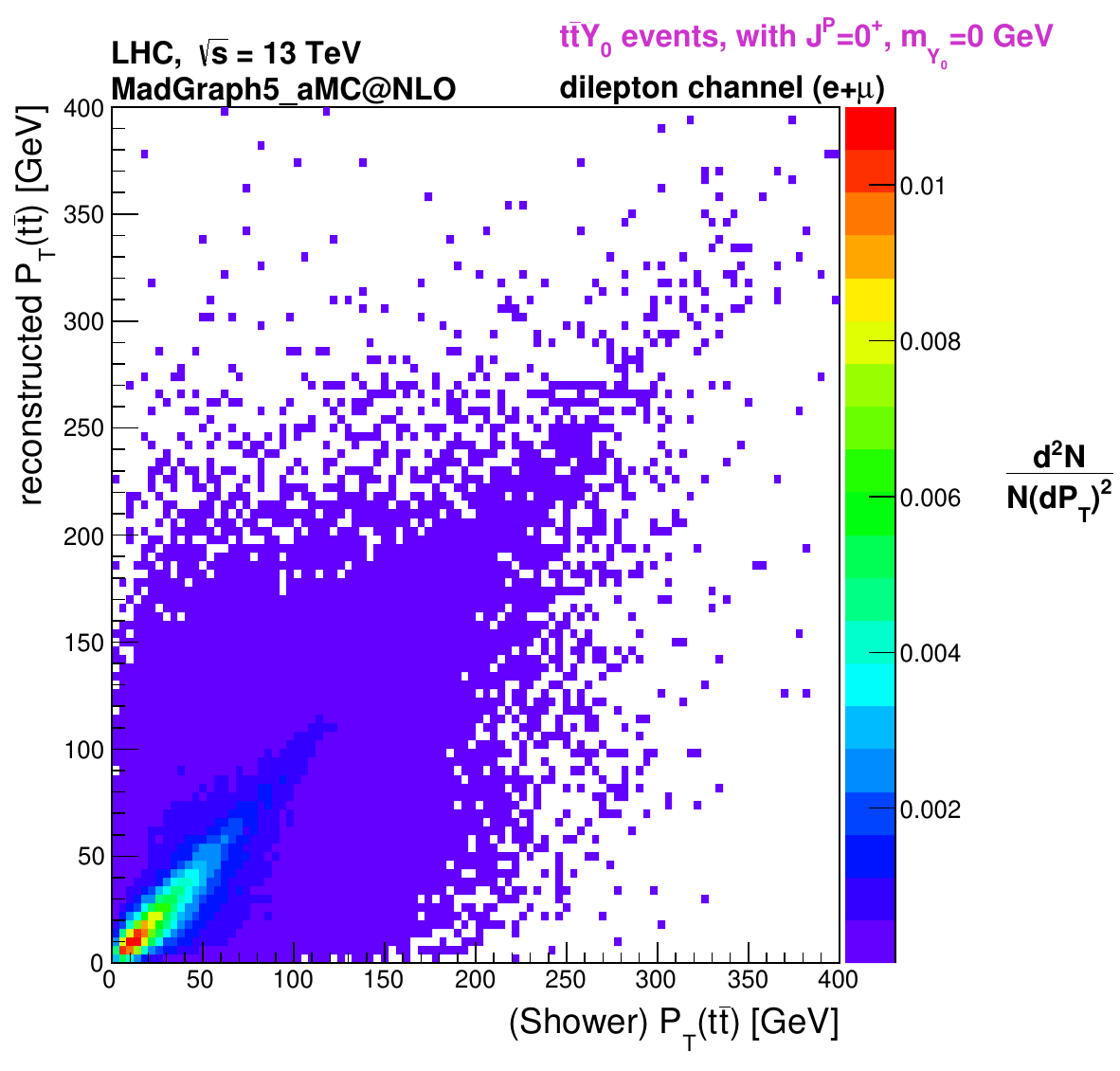}
		\includegraphics[width = 7.0cm]{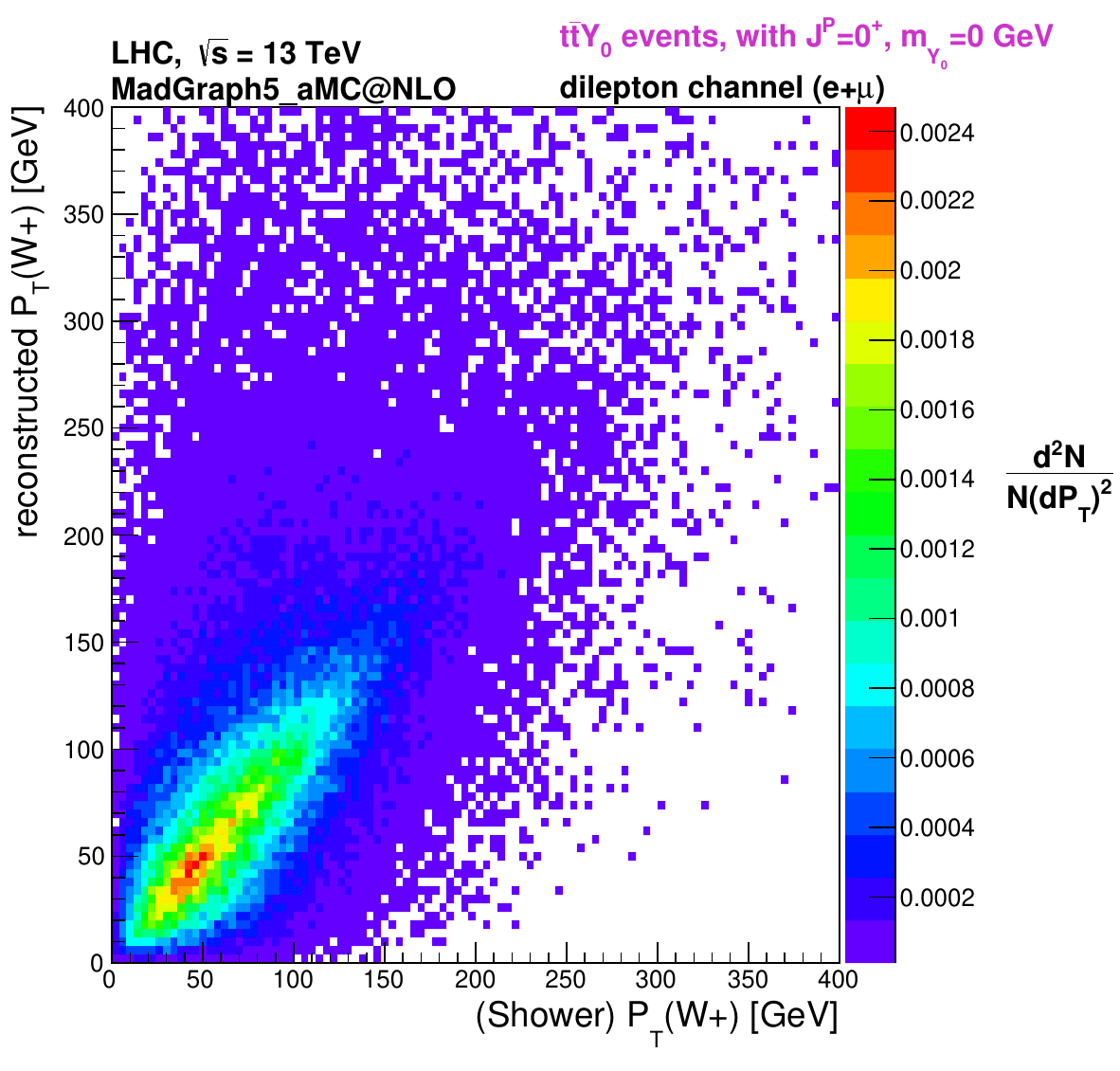} \\
		\caption{Two-Dimensional distributions of $t\bar{t}Y_{0^+}$ events: generator-level transverse momentum versus reconstructed transverse momentum for several particles (neutrino, top left; $t$, top right; $t\overline{t}$ system, bottom left; W$^{+}$ boson, bottom right), for the massless DM mediator.}
		\label{fig:GENvsREC}
	\end{center}
\end{figure} 
The neutrinos four-momenta are obtained from solving the set of equations~\ref{eq:etmiss}, above. If a solution is found, new mass values are tried around the found value (up to 6 times), to check if neighbour mass points can indeed provide a better fit overall. Due to the quadratic form of the mass equations, multiple solutions may exist for a single event. 
We chose the solution that provides the highest value of the likelihood ($L$) constructed using parton level (with shower effects) distributions, in particular the \textit{p.d.f.}s for the transverse momenta of the neutrinos, top quarks and $t\bar{t}$ system, $P(p_{T_{\nu}})$, $P(p_{T_{\bar{\nu}}})$, $P(p_{T_t})$, $P(p_{T_{\bar{t}}})$ and $P(p_{T_{t\bar{t}}})$, respectively. This likelihood is defined according to
\begin{equation}
    L \propto \frac{1}{p_{T_{\nu}}p_{T_{\bar{\nu}}}} P(p_{T_{\nu}})P(p_{T_{\bar{\nu}}})P(p_{T_{t}})P(p_{T_{\bar{t}}})P(p_{T_{t\bar{t}}}) ,
\end{equation}
where the normalization factor ${1}/{p_{T_{\nu}}p_{T_{\bar{\nu}}}}$ is applied to account for the energy losses due to radiation emission and effects from detector resolutions which will increase the reconstructed neutrino 4-momentum. This factor compensates for values of neutrino $p_T$ which are too extreme by giving less weight to those solutions. If no solution is found, the event is discarded. For a scalar (pseudo-scalar) DM mediator mass of 0~GeV, we found that 75\% (72\%) of all events were correctly reconstructed, a number that matches quite well typical SM $t\bar{t}$ analyses where such kinematic reconstruction is attempted.

In Figure~\ref{fig:GENvsREC}, we show the 2-dimensional $p_T$ distributions of the neutrino (top left), top quark (top right), $t\bar{t}$ system (bottom left) and $W^+$ boson (bottom right), after kinematic reconstruction. We can see that the parton level (with shower effects) and the reconstructed kinematics are highly correlated for all particles or systems of particles. This implies that, after experimental effects are taken into account, it is still possible to reconstruct the $t\bar{t}$ system without even trying to reconstruct the invisible DM mediator. This point is quite important since it opens up the possibility of studying the changes in angular distributions of $t\bar{t}$ systems in the presence of a new invisible particle.

\section{Results and Discussion \label{sec:results}}
\hspace{\parindent} 

\begin{figure}[H]
	\begin{center}
		\includegraphics[width = 7.5cm]{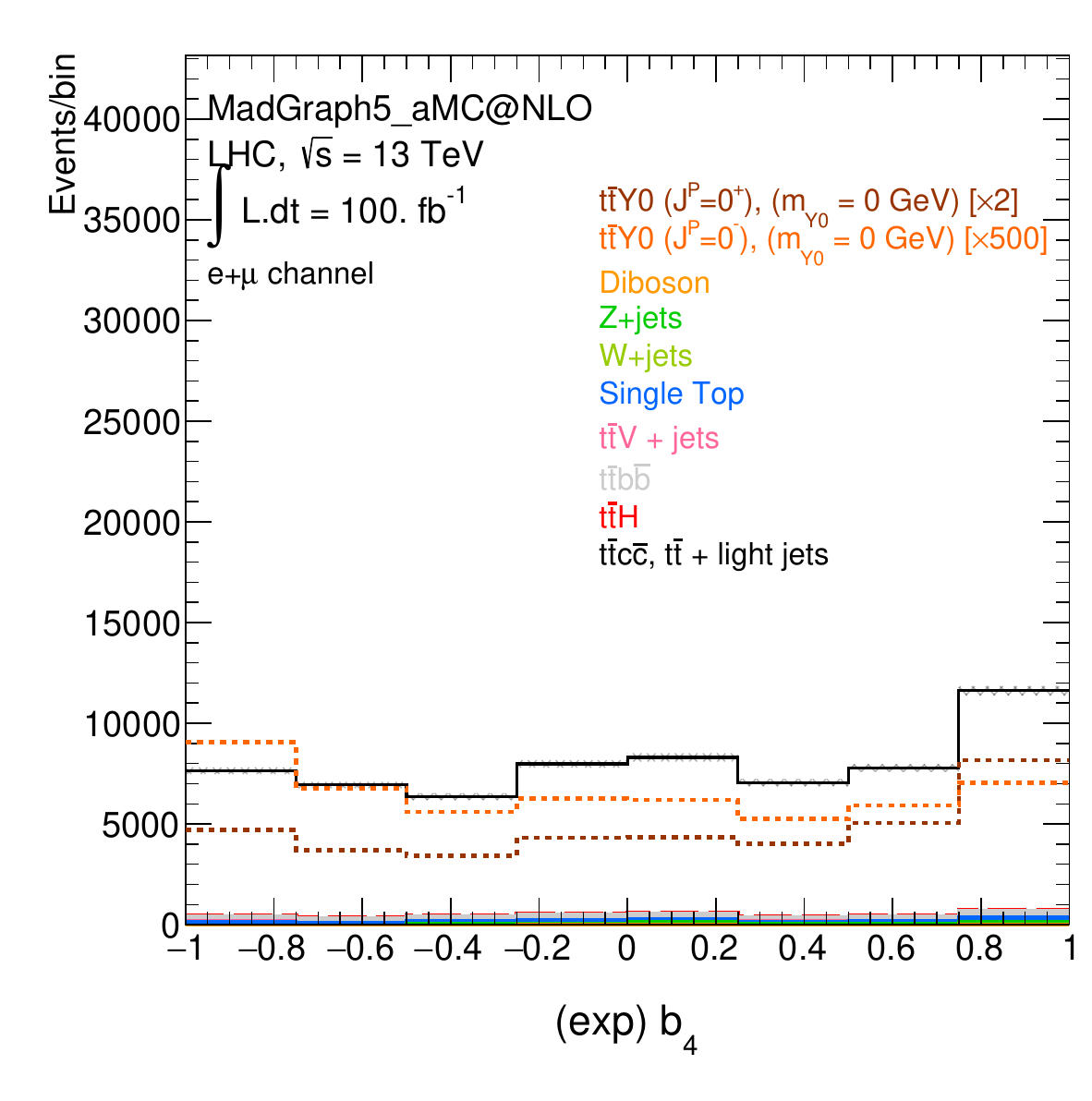}
		\includegraphics[width = 7.5cm]{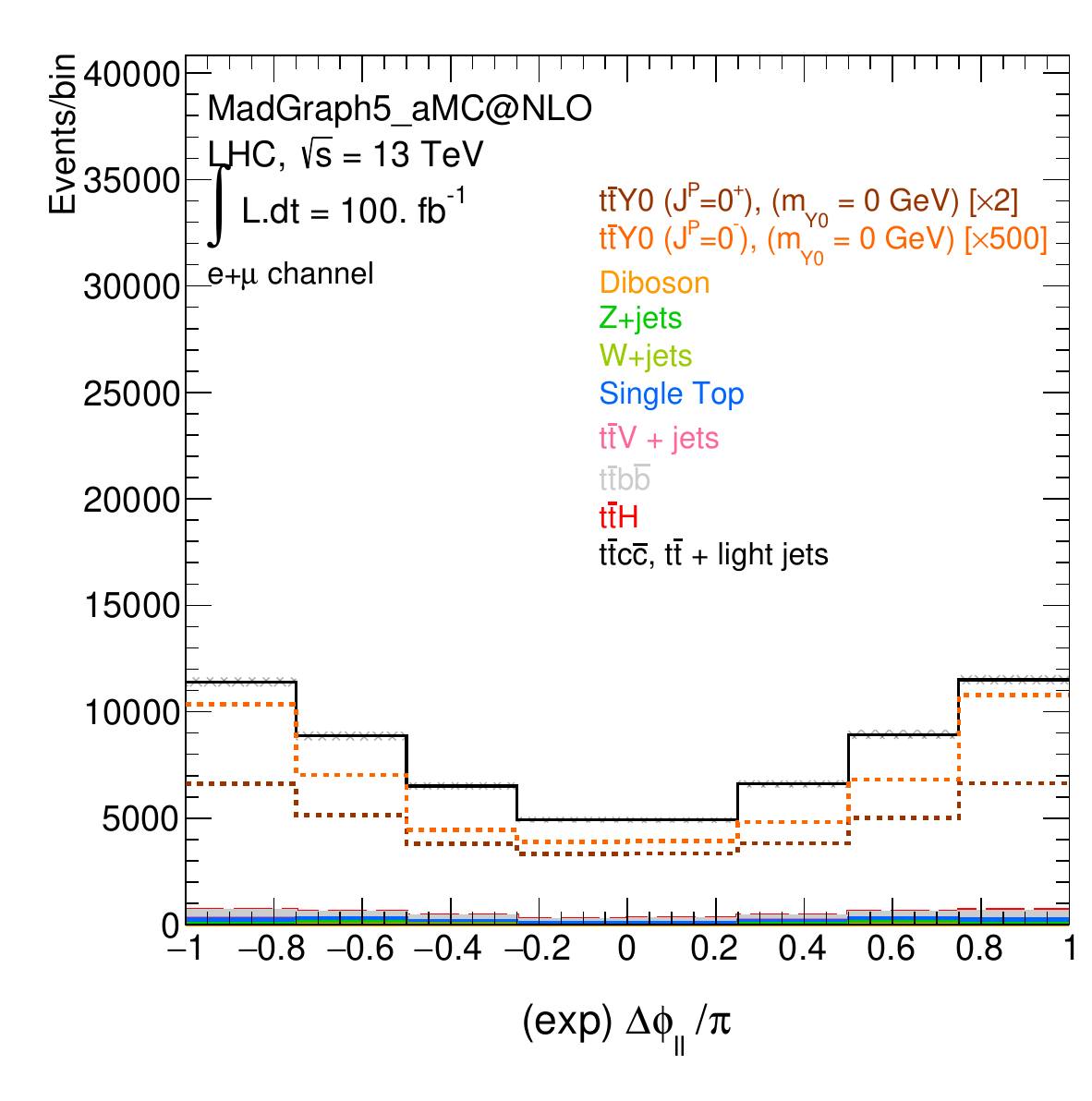}
		\caption{The $b_4$ (left) and $\Delta \phi_{\ell^+ \ell^-}$ (right) distributions for scalar and pseudo-scalar signals (dashed curves) together with the SM processes (full lines) with dileptonic final states, are represented after event selection and kinematic reconstruction (exp), for a reference luminosity of 100~fb$^{-1}$. Scaling factors are applied to the scalar and pseudo-scalar signals for convenience.}
		\label{fig:Stackplots1}
	\end{center}
\end{figure}

\begin{figure}[H]
	\begin{center}
	    \vspace*{1 cm}
		\includegraphics[width = 7.5cm]{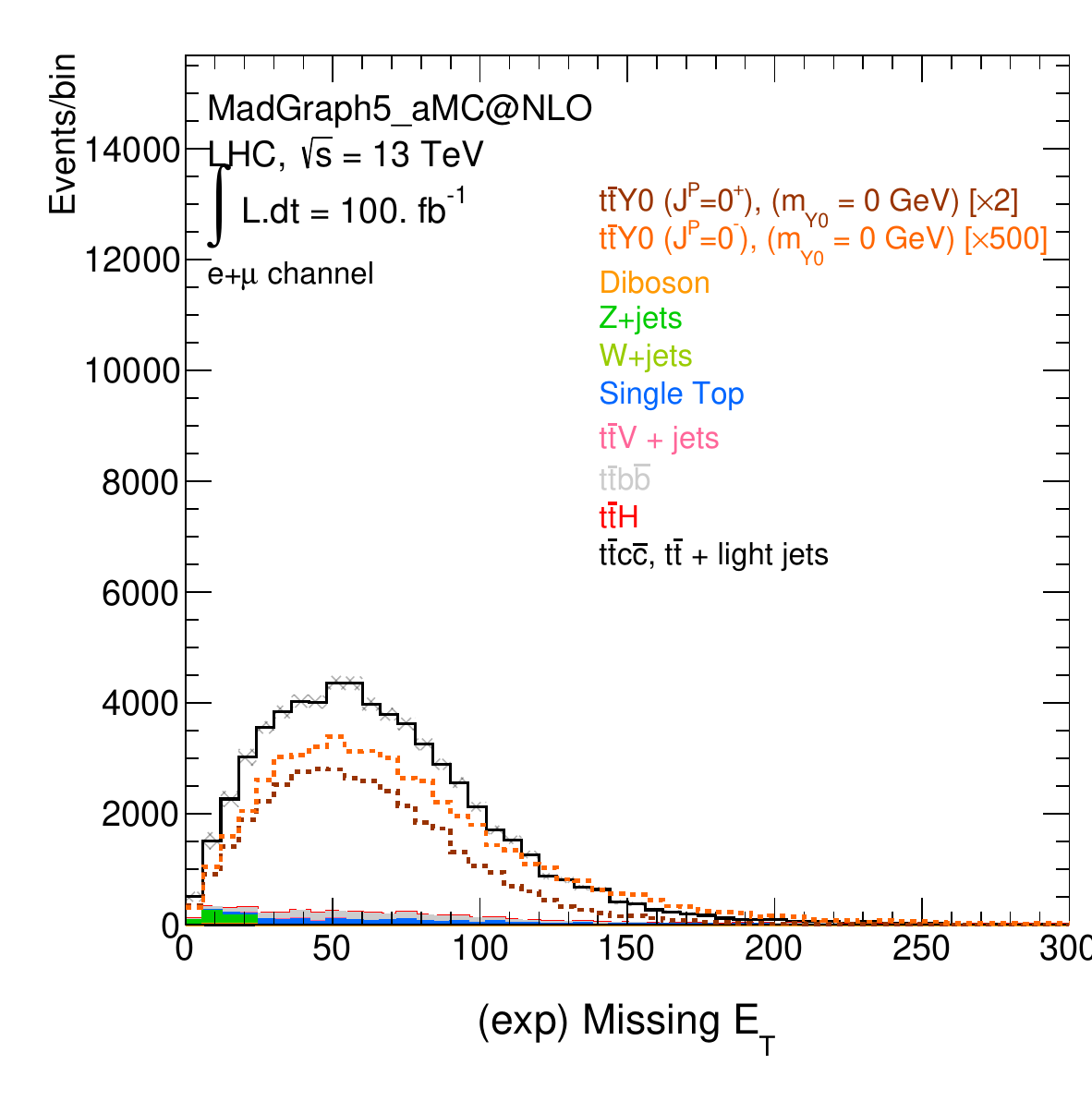}
		\caption{Missing transverse energy ($E_T$) distributions for scalar and pseudo-scalar signals (dashed curves) together with the SM processes (full lines) with dileptonic final states, are represented after event selection and kinematic reconstruction (exp), for a reference luminosity of 100~fb$^{-1}$. Scaling factors are applied to the scalar and pseudo-scalar signals for convenience.}
		\label{fig:Stackplots2}
	\end{center}
\end{figure}

In Figure~\ref{fig:Stackplots1}, we show the $b_4$ (left) and $\Delta \phi_{\ell^+ \ell^-}$ (right) distributions after event selection and kinematic reconstruction, for a reference luminosity of 100~fb$^{-1}$. All SM backgrounds: $t\bar{t}$ ($t\bar{t}c\bar{c}$ and $t\bar{t}$+light jets), $t\bar{t}b\bar{b}$, $t\bar{t}V$, $t\bar{t}H$, single top quark production ($t$-, $s$- and $Wt$-channels), $W/Z$+jets, and diboson ($WW, ZZ, WZ$) events are represented. The $t\bar{t}Y_0$ scalar and pseudo-scalar signals, with $m_{Y_0}=0$~GeV, are shown as well, scaled by factors of 2 and 500 respectively, for convenience. As expected, the main SM background contribution is the $t\bar{t}$ due to its similarity with the signal final state topology. All other backgrounds are essentially residual to the overall SM background contribution. Differences in the shapes of the background distributions can also be noticed when compared with the signals. For instance, in the $b_4$ distribution and for the scalar signal (in brown), events tend to populate positive values more than negative values. This behaviour is inverted for the pseudo-scalar case (in orange). For the $\Delta \phi_{\ell^+ \ell^-}$ distribution (which is symmetric around zero), the scalar and pseudo-scalar signals populate differently its extreme regions. For completeness, we show in Figure~\ref{fig:Stackplots2} the missing transverse energy ($E_T$) distribution, which shows a quite similar behaviour to the SM background one, for both the CP-even and CP-odd signals. This means that missing $E_T$ is not a good discriminating variable for this process which is a very important point to make in an analysis 
for a process with DM in the final state. 

\begin{figure}[H]
        \hspace*{-5mm}\includegraphics[width = 8.2cm]{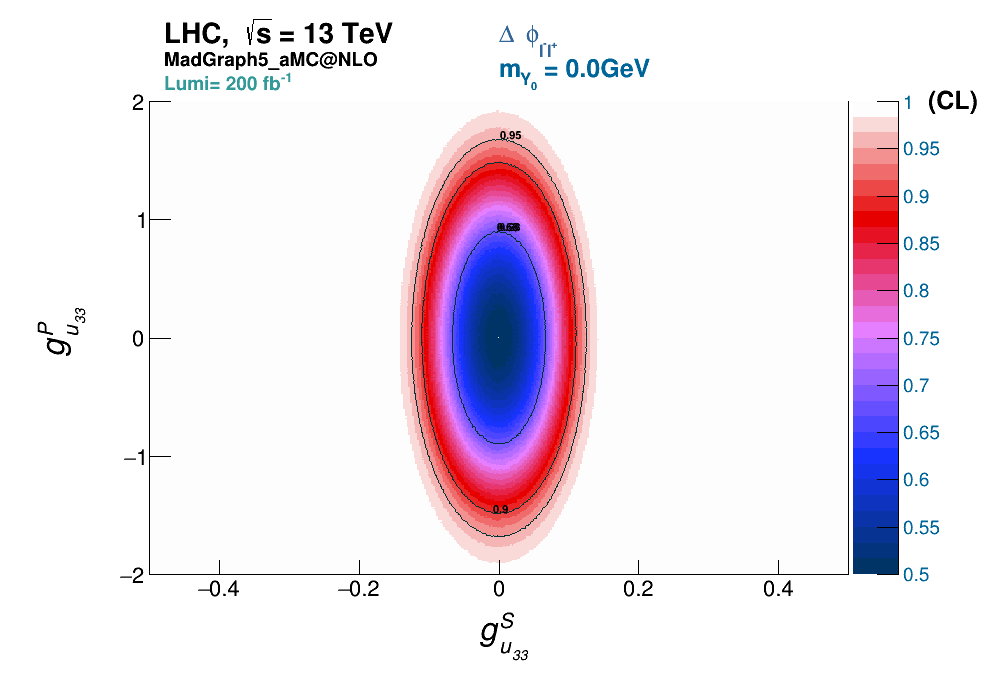}
		\includegraphics[width = 8.2cm]{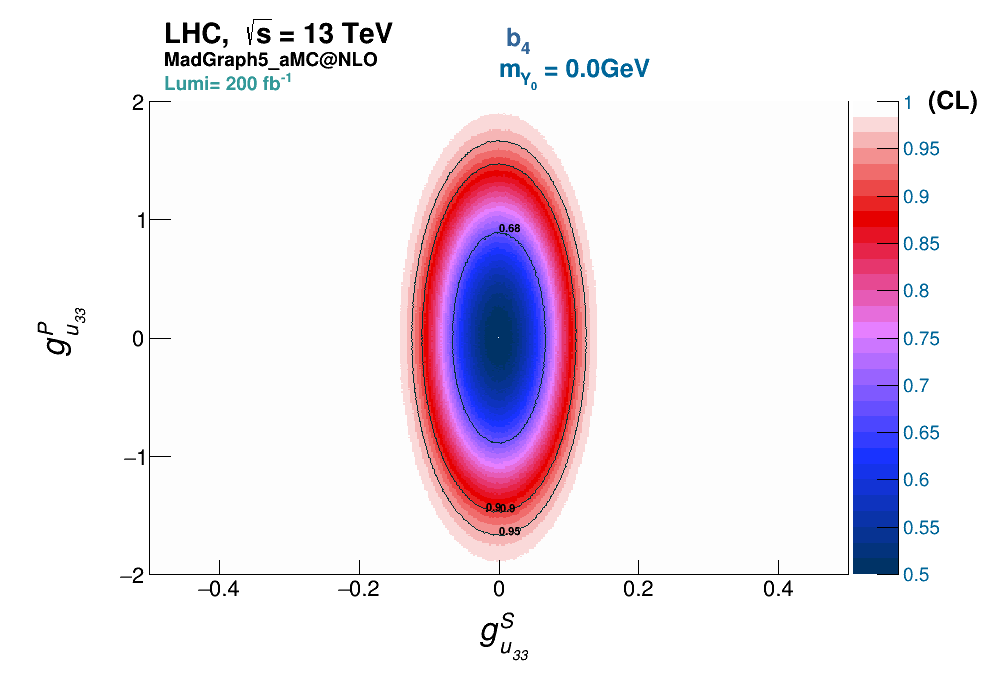}
		\caption{CLs for the exclusion of the SM with a massless DM mediator, $Y_0$, with mixed scalar and pseudo-scalar couplings with the top quarks, against the SM as null hypothesis, for the $\Delta \phi$ between the charged leptons, $\Delta \phi_{\ell^+ \ell^-}$ (left), and $b_4$ (right) observables. Limits are shown for a luminosity of $L=200$~fb$^{-1}$.}
		\label{fig:CLb4vsdeltaphi}
\end{figure}

The $b_4$ and $\Delta \phi_{\ell^+ \ell^-}$ distributions were then used to set confidence level limits (CLs) on the exclusion of the SM with a new CP-mixed massless DM mediator particle, $Y_0$, assuming the SM hypothesis as the null hypothesis (Scenario 1). 
The result is shown in Figure~\ref{fig:CLb4vsdeltaphi}, for an integrated luminosity corresponding roughly to the RUN~2 luminosity plus the contribution from the first year of RUN~3, i.e.,  $L\sim 200$~fb$^{-1}$. 
The CL limits are shown as contour plots in the $(g^S_{u_{33}},g^P_{u_{33}})$ 2D plane. 
It is clear that the CLs are identical for both observables, in this scenario.

The CLs are also evaluated, for Scenario 1, for the full luminosity expected at the end of the High-Luminosisty phase of the LHC (HL-LHC), for $L=3000$~fb$^{-1}$, using the $\Delta \phi_{\ell^+ \ell^-}$ distribution. 
The resulting 68\% and 95\% exclusion limits, for both luminosity values, are in Table~\ref{table:exclusion_limits}. For $L=3000$~fb$^{-1}$, we observe a substantial improvement by factors of 2 to 3, on the exclusion limits. Quite similar results where obtained when using a simple counting experiment. This leads us to conclude that the observable choice has little to no impact on the exclusion limits in this scenario and the DM mediator production cross section is, in itself, the dominant factor. 

\begin{table}[H]
	\renewcommand{\arraystretch}{1.3}
	\begin{center}
		\begin{tabular}{|c|c|cc|cc|}
			
			\hline
			\multicolumn{2}{|c|}{Exclusion Limits}  &    \multicolumn{2}{c|}{$L$ = 200~fb$^{-1}$} & \multicolumn{2}{|c|}{$L$ = 3000~fb$^{-1}$} \\
			
			\multicolumn{2}{|c|}{from $\Delta \phi_{l^+ l^-}$} &  (68\% CL)	& (95\% CL) & (68\% CL) & (95\% CL)  \\ \hline 
			              
			\hspace*{-3mm} \multirow{2}{2.7cm}{\centering $m_{Y_0}$ = 0~GeV} & 	    	          	
			$g^{S}_{u_{33}} \in$ & [-0.067, +0.067]  &   [-0.125, +0.125] & [-0.022, +0.022]  &   [-0.052, +0.052]   \\               
			 & $g^{P}_{u_{33}} \in$ & [-0.91, +0.91] & [-1.71, +1.71]  & [-0.44, +0.44] & [-0.85, +0.85]  
			\\ \hline          
		\end{tabular}
		\caption{Exclusion limits for the $t\bar{t}Y_0$ CP-couplings for fixed luminosities of 200~fb$^{-1}$ and 3000~fb$^{-1}$ of the SM plus $Y_0$, assuming the SM as the null hypothesis.  The limits are shown at confidence levels of 68\% and 95\%, for the $\Delta \phi_{l^+ l^-}$ variable.}
		\label{table:exclusion_limits}
	\end{center}
\end{table}

\begin{figure}[H]
		\hspace*{-5mm}\includegraphics[width = 8.2cm]{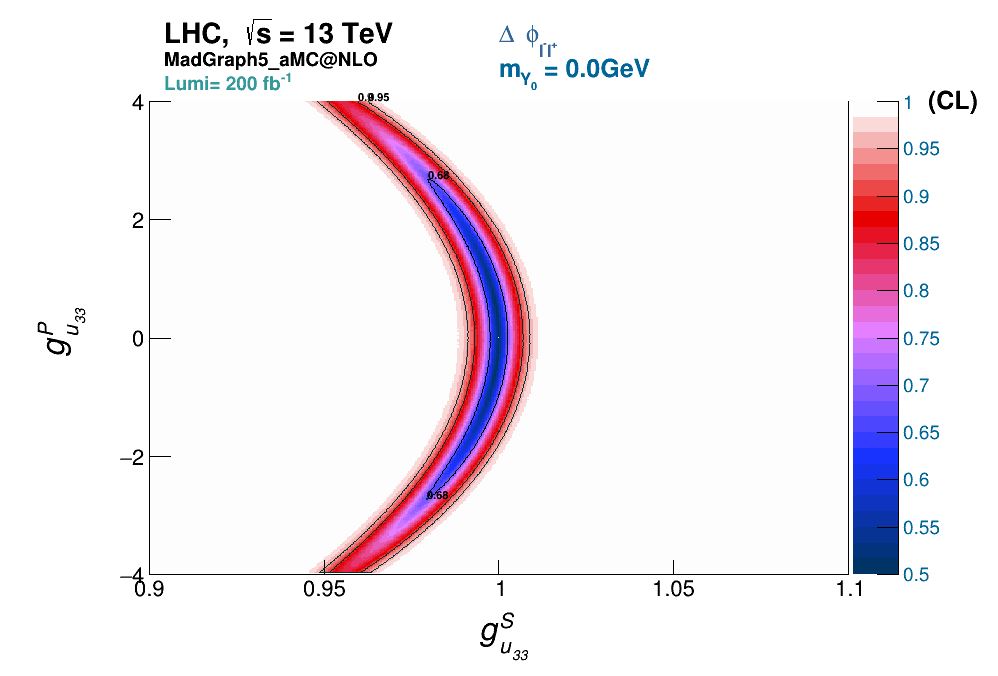}
        \includegraphics[width = 8.2cm]{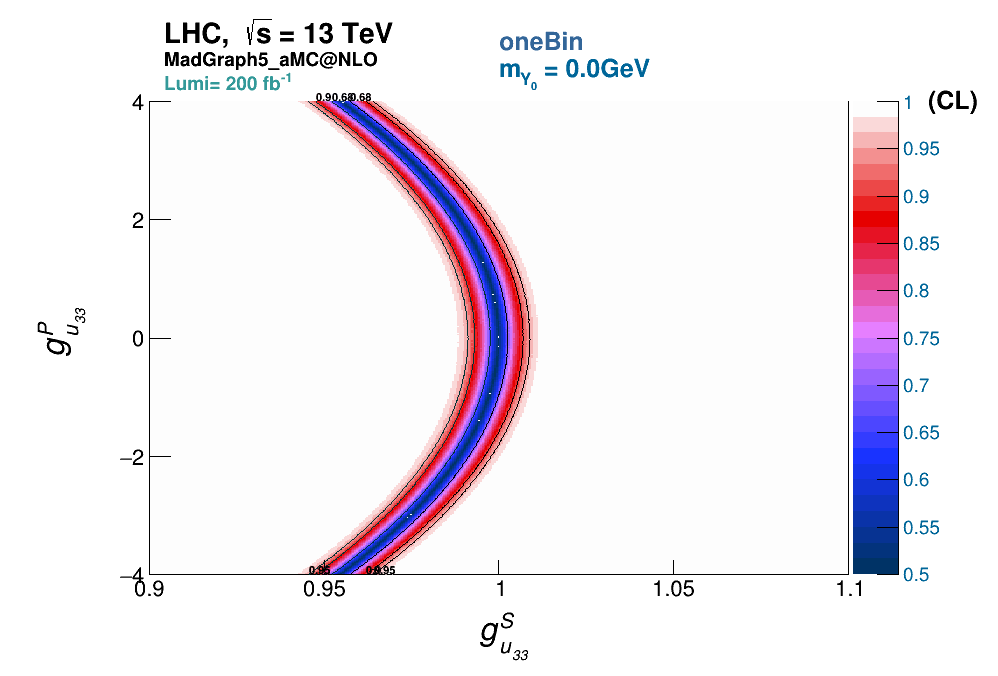}
		\caption{CLs for the exclusion of the SM with a massless DM mediator, $Y_0$, with mixed scalar and pseudo-scalar couplings with the top quarks, against the SM plus a pure scalar DM mediator, for the $\Delta \phi$ between the charged leptons (left). For completeness, the results for a simple event counting experiment (one Bin) is also shown (right). Limits are represented for a luminosity of $L=200$~fb$^{-1}$.}
		\label{fig:onebin}
\end{figure} 

For completeness, an alternative scenario was considered (Scenario 2), where we assumed as null hypothesis the SM plus a pure CP-even DM mediator of mass 0~GeV. The main goal of this scenario is to quantify how well could the mixed state be excluded from a pure CP-even mediator, in case of a discovery. This scenario was explored by using the $\Delta \phi_{\ell^+ \ell^-}$ distribution as well as the simple counting experiment used above for Scenario 1. The results are shown in Figure~\ref{fig:onebin}. Here, however, the difference between both distributions is quite clear, i.e., the 68\% CLs are much worse in the latter case. This indicates that, in Scenario 2, in contrast with Scenario 1, the chosen observable will have an important impact on the exclusion limits. This also means that angular observables can indeed help on studying the CP-nature of DM mediators upon discovery.

\begin{figure}[H]
	\begin{center}
		\hspace{-5mm}
		\includegraphics[height = 5.5cm]{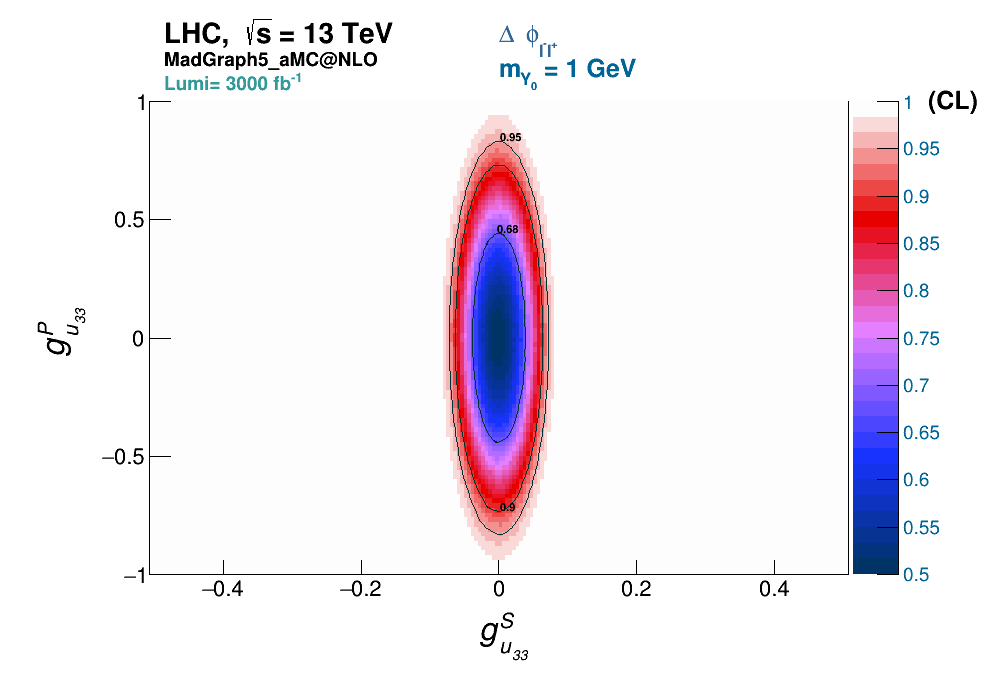}
		\hspace{-5mm}
		\includegraphics[height = 5.5cm]{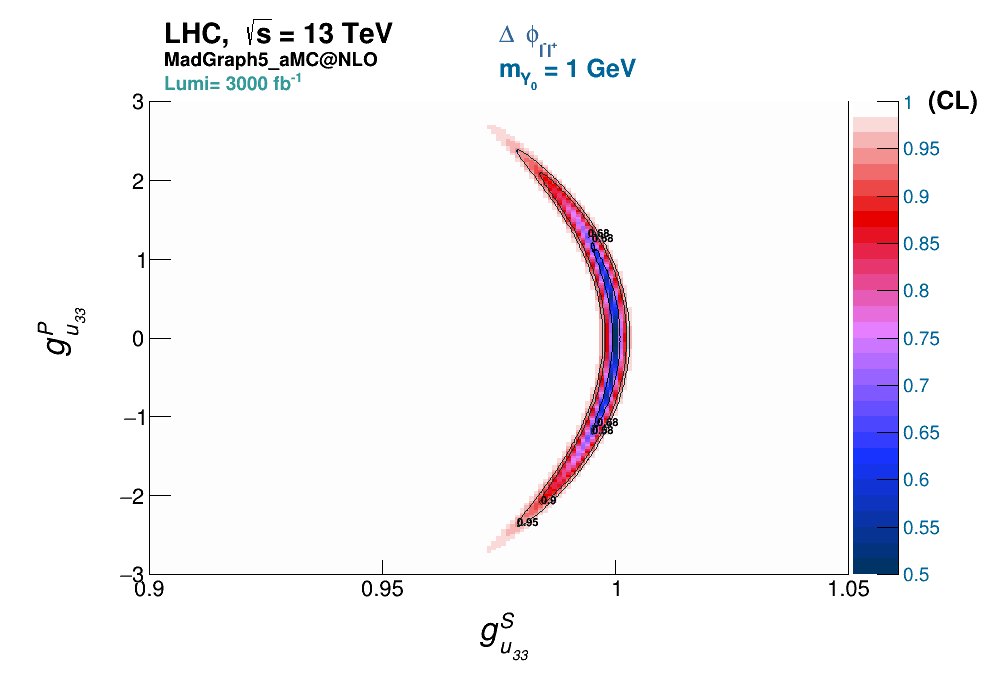} \\
		\includegraphics[height = 5.5cm]{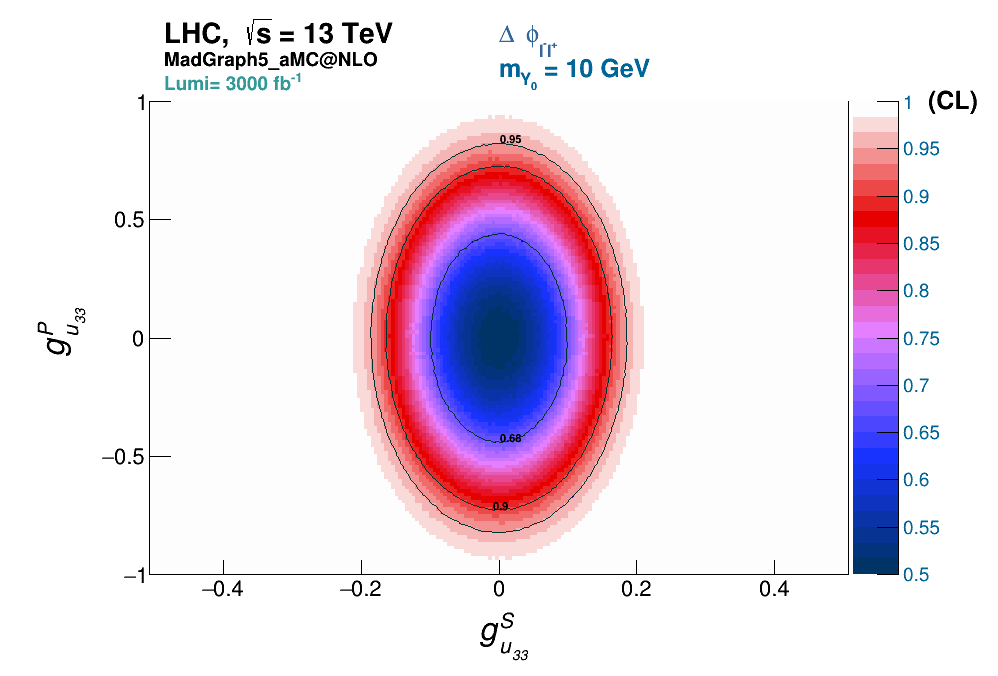}
		\hspace{-5mm}
		\includegraphics[height = 5.5cm]{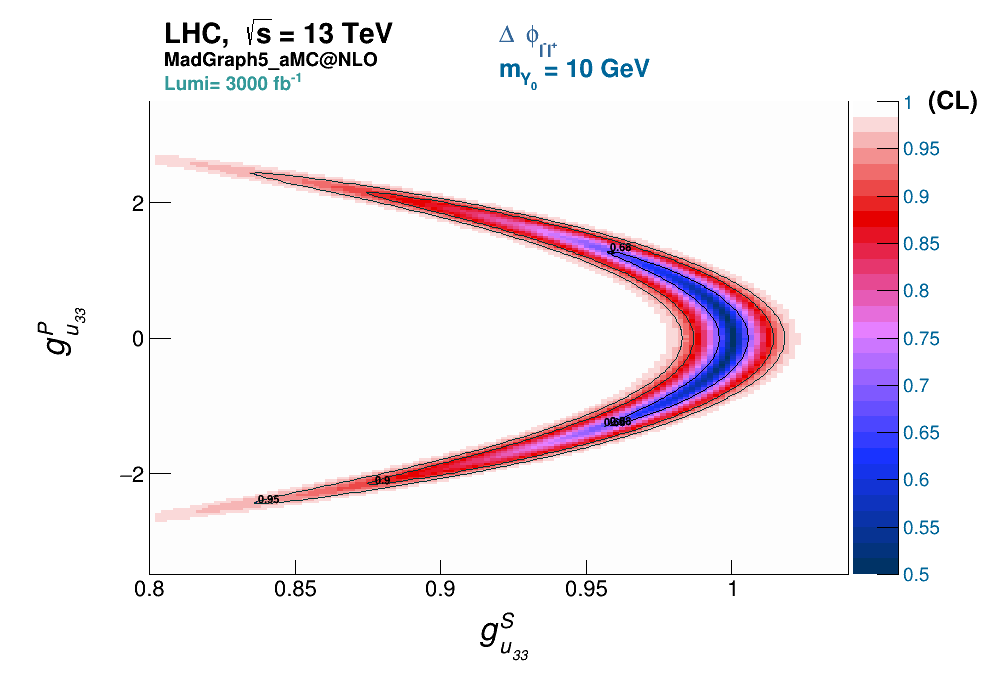} \\
		\includegraphics[height = 5.5cm]{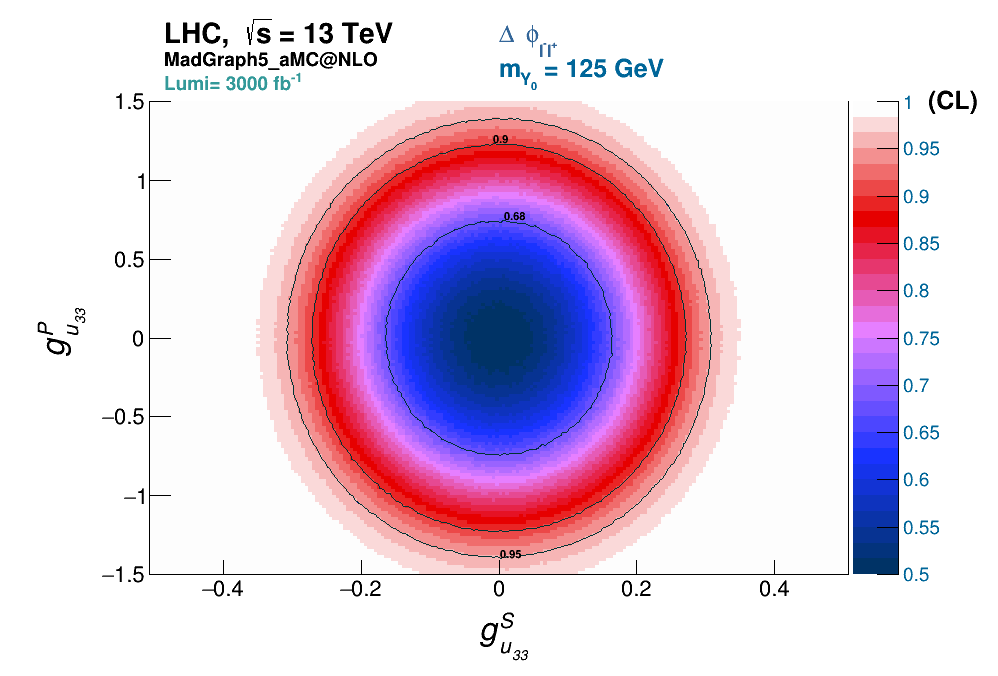}
		\hspace{-5mm}
		\includegraphics[height = 5.5cm]{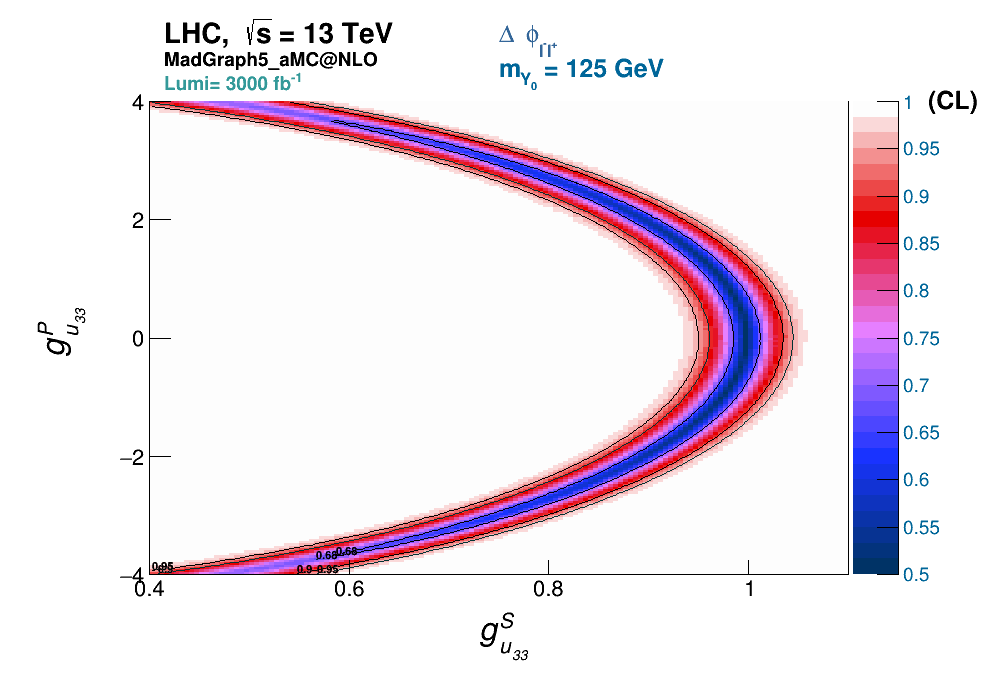}
		\caption{CLs for the exclusion of the SM with a massive DM mediator, $Y_0$ ($m_{Y_0}=$ 1, 10 and 125~GeV in the top, middle, and bottom rows, respectively), with mixed scalar and pseudo-scalar couplings, against the SM as null hypothesis (left), for the $\Delta \phi$ between the charged leptons. For completeness, the corresponding CL for the exclusion of the SM plus a mixed DM mediator against the SM plus a pure scalar DM mediator is also shown for $\Delta \phi$ (right). Limits are shown for a luminosity corresponding to the full HL-LHC luminosity ($L=3000$~fb$^{-1}$).}
		\label{fig:exclusion_limits_masses}
	\end{center}
\end{figure}

Lastly, to extend our results to a massive $Y_0$ produced together with pairs of top quarks, additional signal events were generated with mediator masses ($m_{Y_0}$) set to 1, 10 and 125~GeV. The selection criteria and reconstruction procedure described in Section~\ref{sec:matching} were the same. The resulting $\Delta \phi_{\ell^+ \ell^-}$ distributions were then used to set CLs, in both Scenarios 1 and 2, for a luminosity of 200~fb$^{-1}$ and 3000~fb$^{-1}$ as before. The exclusion limits are depicted for all masses in Figure~\ref{fig:exclusion_limits_masses}, for $L=3000$~fb$^{-1}$. The respective 68\% and 95\% exclusion limits for Scenario 1 are shown in Table~\ref{table:exclusion_limits_masses}, for both $L=200$~fb$^{-1}$ and $L=3000$~fb$^{-1}$. As expected, exclusion limits worsen as masses increase in both Scenarios, since the $t\bar{t}Y_0$ production cross section decreases for heavier $Y_0$ masses, and improve with increasing luminosity values. Also, notice that the observable choice having a very small impact on the exclusion limits in Scenario 1 is true only for smaller DM mediator masses, where the $Y_0$ production cross section is the dominant factor contributing to the exclusion limits.

\begin{table}[H]
	\renewcommand{\arraystretch}{1.3}
	\begin{center}
		\begin{tabular}{|c|c|cc|cc|}
			
			\hline
			\multicolumn{2}{|c|}{Exclusion Limits}  &    \multicolumn{2}{c|}{$L$ = 200~fb$^{-1}$} & \multicolumn{2}{|c|}{$L$ = 3000~fb$^{-1}$} \\
			
			\multicolumn{2}{|c|}{from $\Delta \phi_{l^+ l^-}$} &  (68\% CL)	& (95\% CL) & (68\% CL) & (95\% CL)  \\ \hline 
			              
			\hspace*{-3mm} \multirow{2}{2.7cm}{\centering $m_{Y_0}$ = 1~GeV} & 	    	          	
			$g^{S}_{u_{33}} \in$ & [-0.073, +0.073]  &   [-0.142, +0.142] & [-0.038, +0.038]  &   [-0.068, +0.068]   \\               
			 & $g^{P}_{u_{33}} \in$ & [-0.89, +0.89] & [-1.65, +1.65]  & [-0.43, +0.43] & [-0.83, +0.83]  
			\\ \hline          
            
			\hspace*{-3mm} \multirow{2}{2.7cm}{\centering $m_{Y_0}$ = 10~GeV} & 	    	          	
			$g^{S}_{u_{33}} \in$ & [-0.198, +0.198]  &   [-0.368, +0.372] & [-0.098, +0.098]  &   [-0.188, +0.188]   \\               
			 & $g^{P}_{u_{33}} \in$ & [-0.87, +0.87] & [-1.65, +1.65]  & [-0.44, +0.44] & [-0.83, +0.83]  
			\\ \hline          

           \hspace*{-3mm} \multirow{2}{2.7cm}{\centering $m_{Y_0}$ = 125~GeV} & 	    	          	
			$g^{S}_{u_{33}} \in$ & [-0.328,+0.322]  &   [-0.608, +0.612] & [-0.162, +0.162]  &   [-0.308, +0.308]   \\               
			 & $g^{P}_{u_{33}} \in$ & [-1.48, +1.49] & [-2.77, +2.78]  & [-0.75, +0.75] & [-1.41, +1.41]  
			\\ \hline          
		\end{tabular}
		\caption{Exclusion limits for the $t\bar{t}Y_0$ CP-couplings for fixed luminosities of 200~fb$^{-1}$ and 3000~fb$^{-1}$ of the SM plus $Y_0$, assuming the SM as null hypothesis, for $Y_0$ masses of 1~GeV (top), 10~GeV (middle) and 125~GeV (bottom). The limits are shown at confidence levels of 68\% and 95\%, for the $\Delta \phi_{l^+ l^-}$ variable.}
		\label{table:exclusion_limits_masses}
	\end{center}
\end{table}


\section{Conclusions \label{sec:conclusions}}
\hspace{\parindent} 

In this work we have explored the idea of using on-going searches and measurements, such as the analysis that leads to the measurement of the $t \bar t$ production cross section at the LHC, to look
for hidden DM particles in the final states.
To that end, we present a new approach to fully reconstruct the kinematics of the $t\bar{t}$ system present in $t\bar{t}Y_0$ events produced at the LHC. Our study was performed within the context of simplified models of DM production at colliders. In this kinematic reconstruction, the missing transverse energy was fully attributed to the undetected neutrinos and no attempt to reconstruct the invisible DM mediator was tried. The approximations used in our work appear to be valid in a wider range of the DM mediator mass (starting at $m_{Y_0}=0$~GeV), according to the resulting correlations between the generated and reconstructed kinematics. An example of these correlations is shown in Figure~\ref{fig:GENvsREC} for the $m_{Y_0}=0$~GeV case. Moreover, we have checked that the pairing of the $b$-jets and charged leptons originated from the same parent top quark decay was very well achieved using several angular distributions and dedicated multivariate statistical methods.

We have analyzed a significant number of angular observables, from which two of them were selected to illustrate our findings, the $\Delta \phi_{\ell^+ \ell^-}$ and $b_4$ distributions. These observables were shown to be sensitive not only to DM mediators with different mass scales, but also with different CP-natures, in what concerns their couplings to heavy SM particles. These distributions were then used to set exclusion limits assuming the SM as the null hypothesis, for a luminosity of 200~fb$^{-1}$ and 3000~fb$^{-1}$, corresponding to the full luminosity of the High-Luminosity Phase of the LHC (HL-LHC). We also considered a benchmark scenario that takes into account, as null hypothesis, the SM plus a pure scalar DM mediator, in order to check how sensitive the analysis is to a possible CP-mixed nature of the DM mediator. We observe that, in the former case, the 95\% CL limits using the $\Delta \phi_{\ell^+ \ell^-}$ angular distribution were $g^S_{u_{33}} \in [-0.125, 0.125]$ and $g^P_{u_{33}} \in [-1.71, 1.71]$ for a luminosity of 200~fb$^{-1}$, and $g^S_{u_{33}} \in [-0.052, 0.052]$ and $g^P_{u_{33}} \in [-0.85, 0.85]$ for a luminosity of 3000~fb$^{-1}$. We have also checked that a simple counting experiment can provide similar exclusion limits. In the second case, we observed that the use of angular distributions can improve the exclusion limits for the pseudo-scalar coupling by at least a factor of two, if we want to understand the CP-nature of the DM mediator couplings to SM particles. Finally, we extended our study to the case of a massive DM mediator with $m_{Y_0}$ set to 1, 10 and 125~GeV. We observed that the exclusion limits set from the $\Delta \phi_{\ell^+ \ell^-}$ distributions got worse for increasing $Y_0$ masses, since the $t\bar{t}Y_0$ cross section decreases in that case.

\subsubsection*{Acknowledgments}
%
R.C. and R.S. are partially supported by the Portuguese Foundation for Science and Technology (FCT) under Contracts no. UIDB/00618/2020, UIDP/00618/2020, CERN/FIS-PAR/0025/2021 
CERN/FIS-PAR/0010/2021 and CERN/FIS-PAR/0021/2021.
R.C. is additionally supported by FCT Grant No. 2020.08221.BD. A.O. is partially supported by FCT, under the Contract CERN/FIS-PAR/0037/2021. D.A. is supported by the Deutsche Forschungsgemeinschaft (DFG, German Research Foundation) under grant 396021762 - TRR 257.


\vspace*{1cm}
\bibliographystyle{h-physrev}
\bibliography{papernovo.bib}

\end{document}